\documentclass [] {aa} 
\usepackage{epsfig, graphicx,longtable}
\usepackage{natbib}
\usepackage{times}

\bibpunct{(}{)}{;}{a}{}{,}

\begin{document}


\title{Are beryllium abundances anomalous in stars with giant planets?\thanks{Based 
    on observations collected with the VLT/UT2 Kueyen telescope 
    (Paranal Observatory, ESO, Chile) using the UVES spectrograph 
    (Observing runs 66.C-0116\,A, 66.D-0284\,A, and 68.C-0058\,A), and 
    with the William Herschel and Nordic Optical Telescopes, operated on 
    the island of La Palma by the Isaac Newton Group and jointly by Denmark,
    Finland, Iceland, and Norway, respectively, in the Spanish Observatorio 
    del Roque de los Muchachos of the Instituto de Astrof\'{\i}sica 
    de Canarias.}}


\author{N.C.~Santos\inst{1,2} \and 
        G.~Israelian\inst{3} \and 
        R.J.~Garc\'{\i}a~L\'opez\inst{3,4} \and
	M.~Mayor\inst{2} \and 
	R.~Rebolo\inst{3,5} \and
	S.~Randich\inst{6} \and
        A.~Ecuvillon\inst{3} \and
        C.~Dom\'{\i}nguez~Cerde\~na\inst{3}
	}

\offprints{Nuno C. Santos, \email{Nuno.Santos@oal.ul.pt}}

\institute{
        Centro de Astronomia e Astrof{\'\i}sica da Universidade de Lisboa,
        Observat\'orio Astron\'omico de Lisboa, Tapada da Ajuda, 1349-018
        Lisboa, Portugal
     \and
	Observatoire de Gen\`eve, 51 ch.  des 
	Maillettes, CH--1290 Sauverny, Switzerland
     \and
	Instituto de Astrof{\'\i}sica de Canarias, E-38200 
        La Laguna, Tenerife, Spain
     \and	
        Departamento de Astrof\'{\i}sica, Universidad de La Laguna,
        Av. Astrof\'{\i}sico Francisco S\'anchez s/n, E-38206, La
        Laguna, Tenerife, Spain 
     \and
        Consejo Superior de Investigaciones Cient\'{\i}ficas, Spain 
     \and
        INAF/Osservatorio Astrofisico di Arcetri, Largo Fermi 5, I-50125
        Firenze, Italy}
	
\date{Received / Accepted } 

\titlerunning{Beryllium in planet-host stars} 


\abstract{
In this paper we present beryllium (Be) abundances in a large sample of 41 extra-solar 
planet host stars, and for 29 stars without any known planetary-mass companion,
spanning a large range of effective temperatures. The Be abundances were derived 
through spectral synthesis done in standard Local Thermodynamic Equilibrium,
using spectra obtained with various instruments. The results seem to confirm 
that overall, planet-host stars have ``normal'' Be abundances, although
a small, but not significant, difference might be present. This result is discussed, 
and we show that this difference is probably not due
to any stellar ``pollution'' events. In other words, our results support the idea that the high-metal content
of planet-host stars has, overall, a ``primordial'' origin. However, we also
find a small subset of planet-host late-F and early-G dwarfs that might have 
higher than average Be abundances. The reason for the offset is not clear,
and might be related either to the engulfment of planetary material, to
galactic chemical evolution effects, or to stellar-mass differences for stars
of similar temperature.
\keywords{stars: abundances -- 
          stars: fundamental parameters --
          planetary systems --
	  planetary systems: formation 
          }
}

\maketitle

\section{Introduction}

The study of the chemical abundances in planet-host stars 
\citep[e.g.][]{Gon98,San01,Gon01,Rei02,San03,Law03,Israelian03,San04} has revealed the
important role that the global metallicity plays in the formation of
giant planets. It has been shown that the probability of finding 
a planet is a steeply rising function of the metal content of the 
star \citep[e.g.][]{San01,Rei02,San03,San04}. This crucial observation 
is helping us to better understand the mechanisms involved in the formation of the planetary 
systems \citep[for a general review on extra-solar planets see e.g. ][]{sydney03}. 

Although most studies of the chemical abundances of stars hosting giant
planets have concentrated on the analysis of iron and other iron-peak elements, 
alpha-elements, and other metals \citep[e.g.][]{San00,Gon00,Smi01,Sad02,Bod03,Ecu04,San04}, a 
few have also explored the abundances of the light elements 
$^6$Li \citep[][]{Isr01,Red03,Isr03}, $^7$Li \citep[][]{Gon00,Rya00,Isr04}, and 
$^9$Be \citep[][]{Gar98,Del00,San02}. Overall, and putting aside a few exceptions \citep[e.g.][]{Isr01,Law01}, 
these studies suggest that stars with planets have in general normal light-element abundances, typical
of field stars, even though a few interesting correlations have been 
found \citep[e.g.][]{Isr04}.

\begin{table*}[th]
\caption[]{Spectrographs used for the current study, their spectral resolution, and date of the
observations.}
\begin{tabular}{lccc}
\hline
Spectrograph/Telescope               & Resolution                  & date of       & Designation \\
                                     & ($\lambda$/$\Delta\lambda$) & observations  &             \\
\hline
UES/4.2-m William Herschel Telescope & 55\,000  &  Aug.1998          & UES     \\
IACUB/2.6-m Nordic Optical Telescope & 35\,000  &  May 2000          & IACUB(A)\\  
UVES/VLT 8.2-m Kueyen UT2 (VLT)      & 70\,000  &  Nov.2000-Jan.2001 & UVES(A) \\
UVES/VLT 8.2-m Kueyen UT2 (VLT)      & 56\,000  &  Feb.2001          & UVES(B) \\
IACUB/2.6-m Nordic Optical Telescope & 35\,000  &  Oct.2001          & IACUB(B)\\  
UVES/VLT 8.2-m Kueyen UT2 (VLT)      & 70\,000  &  Oct.2001-Mar.2002 & UVES(C) \\
\hline
\end{tabular}
\label{tabinstr}
\end{table*}

The study of the light elements has an enormous potential for the understanding
of planet formation. First, it is well known that light elements and their 
abundance ratios are good tracers of stellar internal mixing and 
rotation \citep[e.g.][]{Pin90,Ste97}. From the several mixing mechanisms that have been referred
to in the literature as responsible for the depletion of light-elements 
in solar-type stars, rotation and angular momentum loss are among the leading 
processes \citep[see discussion in][]{Ste97}. The study of light element abundances may 
thus probably tell us much about processes related to the angular momentum evolution of 
planet-host stars. If the formation of giant planets needs the
presence of massive proto-planetary disks, we can eventually expect that
planet hosts and ``single'' stars might have had a different angular momentum history,
thus presenting different light-element abundances. 
Although not completely established, a relation between the disk mass
and the rotation history of a star might indeed exist 
\citep[e.g.][]{Edw93,Str94,Sta99,Bar01,Reb01,Reb02,Har02,Wol04}. There is even some theoretical evidence
suggesting that disk life-times might be related to the formation and presence of 
planets \citep[][]{Goo01,Sar04}.
Finally, an angular-momentum variation can also be induced by the eventual 
accretion of planetary-mass bodies into the star \citep[e.g.][]{Sie99,Isr03}.

But other mechanisms exist capable of inducing differences in the light-element abundances 
between planet-host stars and stars without planets. In particular, if at least part of the 
metal ``enrichment'' found for planet host stars is due
to stellar pollution effects \citep[e.g.][]{Lau97,Mur02,Vau04}, we should
also be able to observe an enhancement in the abundances of the light elements in 
planet hosts \citep[e.g.][]{Isr01,Pin01,San02,Isr03}. This enhancement should be at
least of the same order of magnitude as the excess metallicity observed,
although subsequent Li and Be depletion could mask the ``pollution'' effect.

In this paper we continue our study of beryllium (Be) abundances
already started in \citet[][]{San02}, by increasing the number of stars in our samples with
new data taken with the UVES spectrograph (at the VLT/UT2 Kueyen telescope). 
The Be abundances are further compared with Li abundances of the same targets. 
The analysis reveals that overall, and except for a few cases, there are no clear 
differences between planet hosts and stars without any known planetary companion. 
The implications of this result are discussed.
Our analysis also suggests that Be depletion for stars
of different effective temperatures does not behave as expected from models,
and that there seems to exist a Be-gap for solar-temperature stars.
These results are discussed in a separate paper \citep[][]{paperb} -- hereafter, Paper\,B. 


\section{The data}
\label{sec:data}

Part of the spectra analyzed in this paper has already been used in
\citet[][]{San02}. Meanwhile, however, we have gathered near-UV
spectra for more targets using the UVES spectrograph at the 8.2-m
Kueyen VLT (UT2) telescope (run ID 68.C-0058\,A). These new spectra have
a spectral resolution R$\sim$70\,000, and S/N ratios usually between
100 and 200. For a more complete description of
the data obtained with other instruments used we refer to \citet[][]{San02}.

In Table\,\ref{tabinstr} we describe the instruments and 
telescopes used to obtain our data, as well as 
the spectral resolution and the date of the observations. In Tables\,\ref{tabplanets} 
and \ref{tabcomparison} we list the instrument used to observe each star 
as well as the signal-to-noise achieved.

All the data were reduced using IRAF\footnote{IRAF 
is distributed by National Optical Astronomy Observatories, operated 
by the Association of Universities for Research in Astronomy, Inc., 
under contract with the National Science Foundation, U.S.A.} tools 
in the {\tt echelle} package. Standard background correction, 
flat-field, and extraction procedures were used. For the UVES(A),
UVES(C), and UES runs the wavelength calibration was 
done using a ThAr lamp spectrum taken during the same night. For 
the UVES(B), IACUB(A), and IACUB(B) runs the wavelength calibration was 
done using photospheric lines in the region of interest.  

\section{Be abundances}
\label{sec:abundances}

\subsection{Stellar parameters}
\label{sec:parameters}

The stellar atmospheric parameters for most of our targets were taken from
\citet[][]{San04}, who have obtained accurate and uniform stellar
parameters for 98 extra-solar planet-host stars, as well
as for our comparison sample of objects. The errors in the different 
stellar parameters are of the order of 50\,K in T$_{\mathrm{eff}}$, 
0.12\,dex in $\log{g}$, 0.10\,km\,s$^{-1}$ in the microturbulence parameter, 
and 0.05\,dex in [Fe/H]). Only for \object{BD\,-10\,3166}
were the stellar parameters taken from \citet[][]{Gon01}, who
have used a similar technique to derive them. This procedure
gives us the guarantee that the current analysis has no systematic 
errors due to stellar parameter determination (e.g. due to different
effective temperature scales). 

\begin{table*}
\caption[]{Derived Be abundances for the planet hosts stars in our study.}
\begin{tabular}{lcccrrcccccr}
\hline
Star     & T$_{\mathrm{eff}}$ & $\log{g}_{spec}$ & $\xi_{\mathrm{t}}$ & \multicolumn{1}{c}{[Fe/H]} & $\log{N(Be)}$ & $\sigma(Be)$ & Instr.$^\dagger$ & S/N & $v\,\sin{i}$ & source$^{\dagger\dagger}$ & $\log{N(Li)}$\\
         & [K]                &  [cm\,s$^{-2}$]  &  [km\,s$^{-1}$]    &                            &               &              &        &     & [km\,s$^{-1}$]&       &              \\
\hline
\object{BD\,-10\,3166}	&5320	&4.38	 &0.85    &0.33    &$<$ 0.50  &--      &[2]     &20      &1.58	&(a)	 &--    \\
\object{HD\,  6434}	&5835	&4.60	 &1.53    &-0.52   & 1.08   & 0.10   &[3]     &150     &1.30	&(a)	 &$<$ 0.8 \\
\object{HD\,  9826}	&6212	&4.26	 &1.69    &0.13    & 1.05   & 0.13   &[6]     &120     &9	&(b)	 & 2.55 \\
\object{HD\, 10647}	&6143	&4.48	 &1.40    &-0.03   & 1.19   & 0.10   &[3]     &150     &4.87	&(a)	 & 2.80 \\
\object{HD\, 10697}	&5641	&4.05	 &1.13    &0.14    & 1.31   & 0.13   &[5]     &40      &--	&--	 & 1.96 \\
\object{HD\, 12661}	&5702	&4.33	 &1.05    &0.36    & 1.13   & 0.13   &[5]     &40      &--	&--	 &$<$ 0.98\\
\object{HD\, 13445}	&5163	&4.52	 &0.72    &-0.24   &$<$ 0.40  &--      &[1]     &150     &1.27	&(a)	 &$<$-0.12\\
\object{HD\, 16141}	&5801	&4.22	 &1.34    &0.15    & 1.17   & 0.13   &[1]     &120     &1.95	&(a)	 & 1.11 \\
\object{HD\, 17051}	&6252	&4.61	 &1.18    &0.26    & 1.03   & 0.13   &[1]     &150     &5.38	&(a)	 & 2.66 \\
\object{HD\, 19994}	&6190	&4.19	 &1.54    &0.24    & 0.93   & 0.12   &[3]     &140     &8.10	&(a)	 & 1.99 \\
\object{HD\, 22049}	&5073	&4.43	 &1.05    &-0.13   & 0.80   & 0.13   &[5]     &100     &2.13	&(a)	 &$<$ 0.25\\
\object{HD\, 22049}	&5073	&4.43	 &1.05    &-0.13   & 0.75   & 0.31   &[3]     &200     &2.13	&(a)	 &$<$ 0.25\\
\object{HD\, 22049}	&5073	&4.43	 &1.05    &-0.13   & 0.77   & --   &avg     &--     &2.13	&(a)	 &$<$ 0.25\\
\object{HD\, 27442}	&4825	&3.55	 &1.18    &0.39    &$<$ 0.30  &--      &[3]     &110     &0	&(a)	 &$<$-0.47\\
\object{HD\, 38529}	&5674	&3.94	 &1.38    &0.40    &$<$-0.10  &--      &[2]     &60      &--	&--	 &$<$ 0.61\\
\object{HD\, 46375}	&5268	&4.41	 &0.97    &0.20    &$<$ 0.80  &--      &[3]     &90      &--	&--	 &$<$-0.02\\
\object{HD\, 52265}	&6103	&4.28	 &1.36    &0.23    & 1.25   & 0.11   &[1]     &120     &3.95	&(a)	 & 2.88 \\
\object{HD\, 75289}	&6143	&4.42	 &1.53    &0.28    & 1.38   & 0.10   &[2]     &30      &3.81	&(a)	 & 2.85 \\
\object{HD\, 75289}	&6143	&4.42	 &1.53    &0.28    & 1.33   & 0.12   &[1]     &110     &3.81	&(a)	 & 2.85 \\
\object{HD\, 75289}	&6143	&4.42	 &1.53    &0.28    & 1.36   &--      &avg     &110     &3.81	&(a)	 & 2.85 \\
\object{HD\, 82943}	&6016	&4.46	 &1.13    &0.30    & 1.37   & 0.17   &[4]     &20      &1.65	&(a)	 & 2.51 \\
\object{HD\, 82943}	&6016	&4.46	 &1.13    &0.30    & 1.27   & 0.12   &[2]     &35      &1.65	&(a)	 & 2.51 \\
\object{HD\, 82943}	&6016	&4.46	 &1.13    &0.30    & 1.27   & 0.12   &[1]     &140     &1.65	&(a)	 & 2.51 \\
\object{HD\, 82943}$^{\star}$	&6016	&4.46	 &1.13    &0.30    & 1.27   &--      &avg     &140     &1.65	&(a)	 & 2.51 \\
\object{HD\, 83443}	&5454	&4.33	 &1.08    &0.35    &$<$ 0.70  &--      &[3]     &100     &1.38	&(a)	 &$<$ 0.52\\
\object{HD\, 92788}	&5821	&4.45	 &1.16    &0.32    & 1.19   & 0.11   &[2]     &40      &1.78	&(a)	 & 1.34 \\
\object{HD\, 95128}	&5954	&4.44	 &1.30    &0.06    & 1.23   & 0.11   &[4]     &100     &2.1	&(c)	 & 1.83 \\
\object{HD\,108147}	&6248	&4.49	 &1.35    &0.20    & 0.99   & 0.10   &[2]     &60      &5.34	&(a)	 & 2.33 \\
\object{HD\,114762}	&5884	&4.22	 &1.31    &-0.70   & 0.82   & 0.11   &[4]     &65      &1.5	&(c)	 & 2.20 \\
\object{HD\,117176}	&5560	&4.07	 &1.18    &-0.06   & 0.86   & 0.13   &[4]     &70      &0.5	&(c)	 & 1.88 \\
\object{HD\,120136}	&6339	&4.19	 &1.70    &0.23    &$<$ 0.25  &--      &[6]     &90      &14.5	&(b)	 &--	\\
\object{HD\,121504}	&6075	&4.64	 &1.31    &0.16    & 1.33   & 0.11   &[2]     &45      &2.56	&(a)	 & 2.65 \\
\object{HD\,130322}	&5392	&4.48	 &0.85    &0.03    & 0.95   & 0.13   &[4]     &35      &1.47	&(a)	 &$<$ 0.13\\
\object{HD\,134987}	&5776	&4.36	 &1.09    &0.30    & 1.22   & 0.11   &[2]     &60      &2.22	&(a)	 &$<$ 0.74\\
\object{HD\,143761}	&5853	&4.41	 &1.35    &-0.21   & 1.11   & 0.12   &[6]     &120     &1.5	&(c)	 & 1.46 \\
\object{HD\,145675}	&5311	&4.42	 &0.92    &0.43    &$<$ 0.65  &--      &[4]     &65      &1	&(d)	 &$<$ 0.03\\
\object{HD\,168443}	&5617	&4.22	 &1.21    &0.06    & 1.11   & 0.13   &[4]     &55      &1.68	&(a)	 &$<$ 0.78\\
\object{HD\,169830}	&6299	&4.10	 &1.42    &0.21    &$<$-0.40  &--      &[3]     &130     &3.35	&(a)	 &$<$ 1.16\\
\object{HD\,179949}	&6260	&4.43	 &1.41    &0.22    & 1.08   & 0.10   &[3]     &100     &6.10	&(a)	 & 2.65 \\
\object{HD\,187123}	&5845	&4.42	 &1.10    &0.13    & 1.08   & 0.12   &[4]     &55      &1.73	&(d)	 & 1.21 \\
\object{HD\,192263}	&4947	&4.51	 &0.86    &-0.02   &$<$ 0.90  &--      &[3]     &60      &2.02	&(a)	 &$<$-0.39\\
\object{HD\,195019}	&5842	&4.32	 &1.27    &0.08    & 1.15   & 0.12   &[4]     &50      &1.73	&(a)	 & 1.47 \\
\object{HD\,202206}	&5752	&4.50	 &1.01    &0.35    & 1.04   & 0.11   &[3]     &130     &2.44	&(a)	 & 1.04 \\
\object{HD\,209458}	&6117	&4.48	 &1.40    &0.02    & 1.24   & 0.11   &[3]     &150     &3.65	&(a)	 & 2.70 \\
\object{HD\,210277}	&5532	&4.29	 &1.04    &0.19    & 0.91   & 0.13   &[1]     &110     &1.39	&(a)	 &$<$ 0.30\\
\object{HD\,217014}	&5804	&4.42	 &1.20    &0.20    & 1.02   & 0.12   &[6]     &100     &2.1	&(c)	 & 1.30 \\
\object{HD\,217107}	&5646	&4.31	 &1.06    &0.37    & 0.96   & 0.13   &[1]     &120     &1.37	&(a)	 &$<$ 0.40\\
\object{HD\,222582}	&5843	&4.45	 &1.03    &0.05    & 1.14   & 0.11   &[3]     &125     &1.75	&(a)	 &$<$ 0.59\\
\hline
\end{tabular}
\\ $^\dagger$ The instruments used to obtain the spectra were: [1] UVES(A); [2] UVES(B); [3] UVES(C); [4] IACUB(A); [5] IACUB(B); [6] UES
\newline
$^{\dagger\dagger}$ The sources of the $v\,\sin{i}$ are: (a) CORALIE \citep[][]{San02b}; (b) \citet[][]{Gon97}; (c) \citet[][]{Gon98}; (d) \citet[][]{Nae04}
\newline
$^{\star}$ Given the lower S/N of the IACUB spectrum, only the UVES spectra were considered.
\label{tabplanets}
\end{table*}

For three of the stars (\object{HD\,870}, \object{HD\,1461}, and \object{HD\,3823} 
for which no planetary companions have been found to date), no 
parameters were available, as these stars were not part of the
volume-limited comparison sample studied in \citet[][]{San01,San04}. As already done 
in \citet[][]{San02}, these were derived from CORALIE or FEROS spectra, in the
same way as for all the other stars.

\begin{table*}
\caption[]{Derived Be abundances for the comparison stars in our study.}
\begin{tabular}{lcccrrcccccr}
\hline
Star     & T$_{\mathrm{eff}}$ & $\log{g}_{spec}$ & $\xi_{\mathrm{t}}$ & \multicolumn{1}{c}{[Fe/H]} & $\log{N(Be)}$ & $\sigma(Be)$ & Instr.$^\dagger$ & S/N & $v\,\sin{i}$ & source$^{\dagger\dagger}$ & $\log{N(Li)}$\\
         & [K]                &  [cm\,s$^{-2}$]  &  [km\,s$^{-1}$]    &                            &               &              &        &     & [km\,s$^{-1}$]&       &              \\
\hline
\object{HD\,   870}	&5447	& 4.57    &1.13   & -0.07   & 0.80   & 0.15   &[1]     &130	&1.77	 &(a)	  &$<$ 0.20 \\
\object{HD\,  1461}	&5768	& 4.37    &1.27   & 0.17    & 1.14   & 0.13   &[1]     &120	&1.71	 &(a)	  &$<$ 0.51 \\
\object{HD\,  1581}	&5956	& 4.39    &1.07   & -0.14   & 1.15   & 0.11   &[1]     &140	&2.16	 &(a)	  & 2.37  \\
\object{HD\,  3823}	&5948	& 4.06    &1.17   & -0.25   & 1.02   & 0.12   &[1]     &130	&1.99	 &(a)	  & 2.41  \\
\object{HD\,  4391}	&5878	& 4.74    &1.13   & -0.03   & 0.64   & 0.11   &[3]     &150	&2.72	 &(a)	  &$<$ 1.09 \\
\object{HD\,  7570}	&6140	& 4.39    &1.50   & 0.18    & 1.17   & 0.10   &[3]     &180	&3.82	 &(a)	  & 2.91  \\
\object{HD\, 10700}	&5344	& 4.57    &0.91   & -0.52   & 0.83   & 0.11   &[3]     &180	&0.90	 &(a)	  &$<$ 0.41 \\
\object{HD\, 14412}	&5368	& 4.55    &0.88   & -0.47   & 0.80   & 0.11   &[3]     &190	&1.42	 &(a)	  &$<$ 0.44 \\
\object{HD\, 20010}	&6275	& 4.40    &2.41   & -0.19   & 1.01   & 0.10   &[3]     &180	&4.63	 &(a)	  & 2.13  \\
\object{HD\, 20766}	&5733	& 4.55    &1.09   & -0.21   &$<$-0.09  &--      &[3]     &200	&1.98	 &(a)	  &$<$ 0.97 \\
\object{HD\, 20794}	&5444	& 4.47    &0.98   & -0.38   & 0.91   & 0.11   &[3]     &250	&0.52	 &(a)	  &$<$ 0.52 \\
\object{HD\, 20807}	&5843	& 4.47    &1.17   & -0.23   & 0.36   & 0.11   &[3]     &160	&1.74	 &(a)	  &$<$ 1.07 \\
\object{HD\, 23249}	&5074	& 3.77    &1.08   & 0.13    &$<$ 0.15  &--      &[5]     &80	&1.01	 &(a)	  & 1.24 \\
\object{HD\, 23484}	&5176	& 4.41    &1.03   & 0.06    &$<$ 0.70  &--      &[3]     &140	&2.40	 &(a)	  &$<$ 0.40 \\
\object{HD\, 26965\,A}	&5126	& 4.51    &0.60   & -0.31   & 0.76   & 0.13   &[5]     &55	&0.77	 &(a)	  &$<$ 0.17 \\
\object{HD\, 30495}	&5868	& 4.55    &1.24   & 0.02    & 1.16   & 0.11   &[3]     &140	&3.04	 &(a)	  & 2.44  \\
\object{HD\, 36435}	&5479	& 4.61    &1.12   & 0.00    & 0.99   & 0.12   &[3]     &210	&4.58	 &(a)	  & 1.67  \\
\object{HD\, 38858}	&5752	& 4.53    &1.26   & -0.23   & 1.02   & 0.11   &[3]     &150	&0.99	 &(a)	  & 1.64  \\
\object{HD\, 43162}	&5633	& 4.48    &1.24   & -0.01   & 1.08   & 0.11   &[3]     &160	&5.49	 &(a)	  & 2.34  \\
\object{HD\, 43834}	&5594	& 4.41    &1.05   & 0.10    & 0.94   & 0.11   &[3]     &220	&1.44	 &(a)	  & 2.30  \\
\object{HD\, 69830}	&5410	& 4.38    &0.89   & -0.03   & 0.79   & 0.11   &[3]     &100	&0.75	 &(a)	  &$<$ 0.47 \\
\object{HD\, 72673}	&5242	& 4.50    &0.69   & -0.37   & 0.70   & 0.13   &[3]     &180	&1.19	 &(a)	  &$<$ 0.48 \\
\object{HD\, 74576}	&5000	& 4.55    &1.07   & -0.03   & 0.70   & 0.31   &[3]     &120	&3.56	 &(a)	  & 1.72  \\
\object{HD\, 76151}	&5803	& 4.50    &1.02   & 0.14    & 1.02   & 0.11   &[3]     &110	&1.02	 &(a)	  & 1.88  \\
\object{HD\, 84117}	&6167	& 4.35    &1.42   & -0.03   & 1.11   & 0.11   &[3]     &160	&4.85	 &(a)	  & 2.64  \\
\object{HD\,189567}	&5765	& 4.52    &1.22   & -0.23   & 1.06   & 0.10   &[3]     &160	&1.29	 &(a)	  &$<$ 0.82 \\
\object{HD\,192310}	&5069	& 4.38    &0.79   & -0.01   &$<$ 0.60  &--      &[3]     &180	&0.85	 &(a)	  &$<$ 0.20 \\
\object{HD\,211415}	&5890	& 4.51    &1.12   & -0.17   & 1.12   & 0.10   &[3]     &190	&1.84	 &(a)	  & 1.92  \\
\object{HD\,222335}	&5260	& 4.45	  &0.92   & -0.16   & 0.66   & 0.22   &[1]     &110	&1.25	 &(a)	  &$<$ 0.31 \\
\hline
\end{tabular}
\\ $^\dagger$ The instruments used to obtain the spectra were: [1] UVES(A); [2] UVES(B); [3] UVES(C); [4] IACUB(A); [5] IACUB(B); [6] UES
\newline
$^{\dagger\dagger}$ The sources of the $v\,\sin{i}$ are: (a) CORALIE \citep[][]{San02b}; (b) \citet[][]{Gon97}; (c) \citet[][]{Gon98}; (d) \citet[][]{Nae04}
\label{tabcomparison}
\end{table*}

\begin{figure*}[t]
\psfig{width=\hsize,file=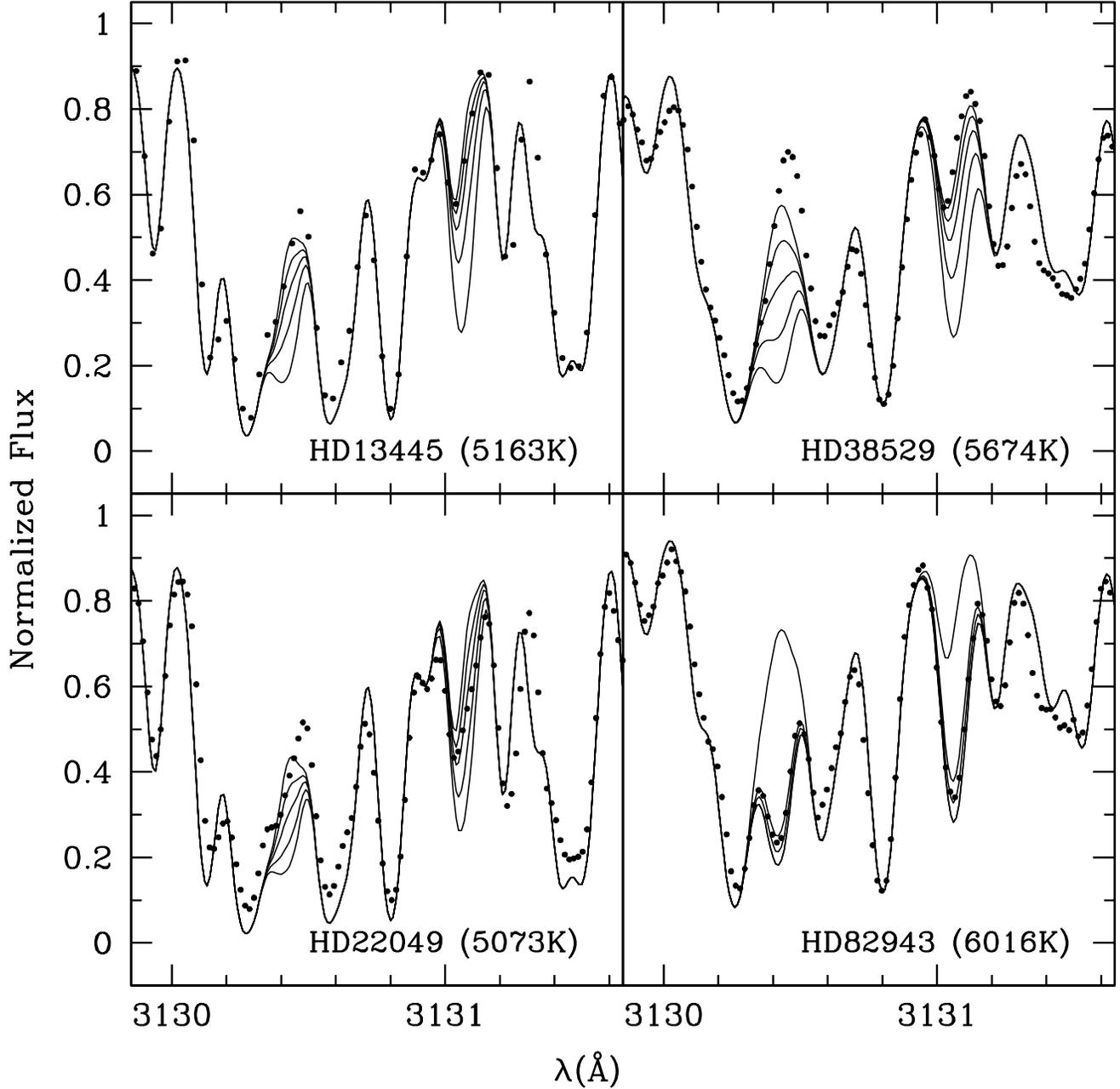}
\caption[]{Spectral syntheses (lines) and observed spectra (dots) in 
the \ion{Be}{ii} line region for 4 extra-solar planet-host stars of different effective 
temperatures (see Table\,\ref{tabplanets}). In all panels, the upper and lower 
syntheses were done with a $\log{N(Be)}$ of 1.42 (meteoritic) and $-$10.0 (essentially no Be).
{\it Lower panels}: For \object{HD\,82943}, the three other spectral syntheses correspond to the optimal fit and to fits with 
abundance variations of $\pm$0.15\,dex. For \object{HD\,22049}, the three intermediate fits were done with Be abundances 
of 1.05, 0.75 and 0.25\,dex. {\it Upper panels}: fits for two stars for which we 
have obtained only upper limits for the Be abundances. \object{HD\,13445}: the syntheses 
correspond to Be abundances of 1.42, 0.90, 0.50, 0.10, and $-$10.0; \object{HD\,38529}: the 
syntheses correspond to Be abundances of 1.42, 0.80, 0.40, $-$0.10, and $-$10.0. Stellar effective temperatures
are also shown.}
\label{figplanets}
\end{figure*}

\begin{figure*}[t]
\psfig{width=\hsize,file=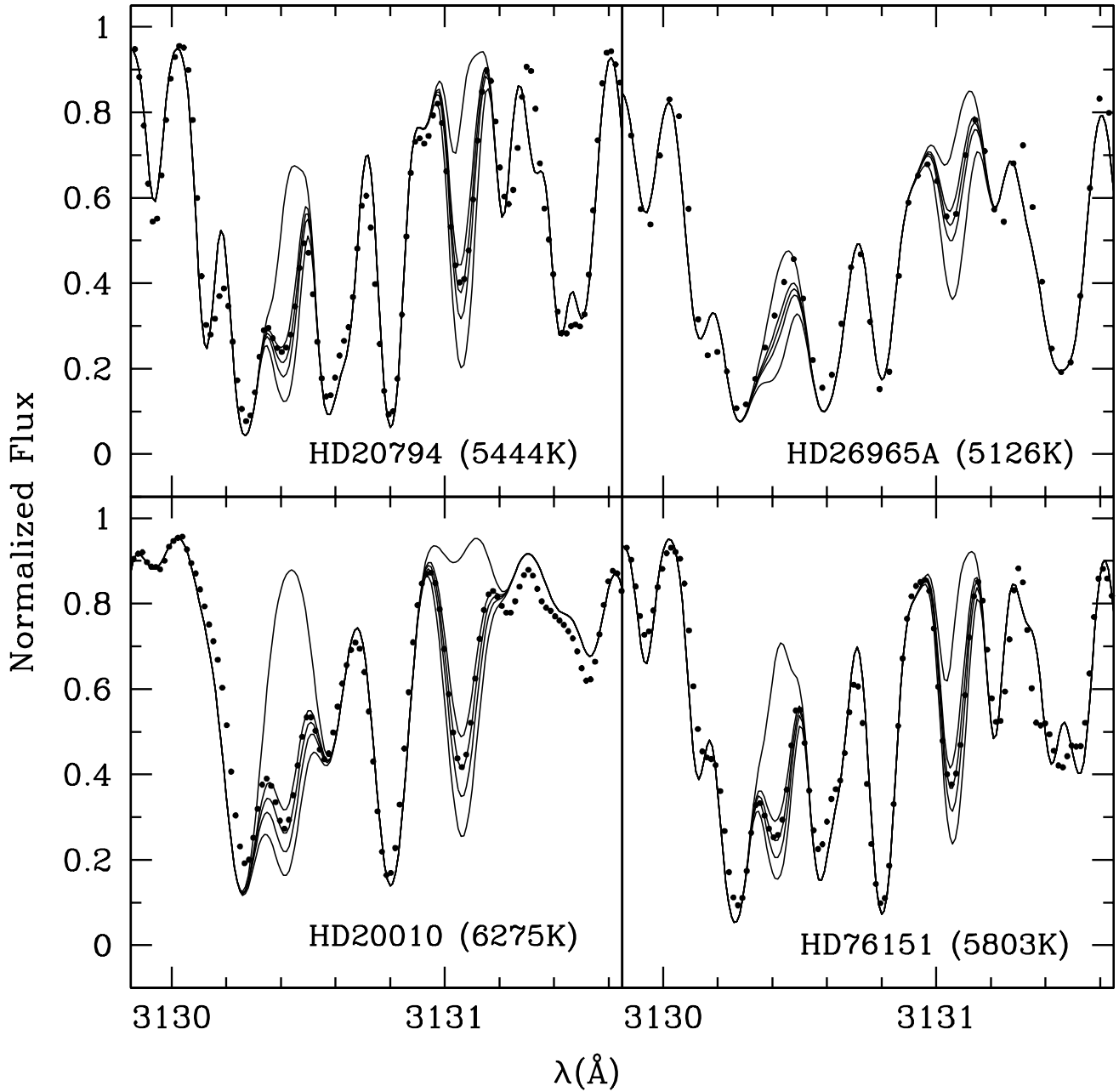}
\caption[]{Same as Fig.\,\ref{figplanets} but for 4 stars for which no planetary companions have
been detected. In all panels we present the observed data and
synthetic spectra obtained with meteoritic and ``no-Be'' (upper and lower fits, respectively), together 
with three other syntheses. These correspond to the optimal fit and to fits with 
abundance variations of $\pm$0.15\,dex. Stellar effective temperatures
are also shown.}
\label{figcomparison}
\end{figure*}

\begin{figure}[t]
\psfig{width=\hsize,file=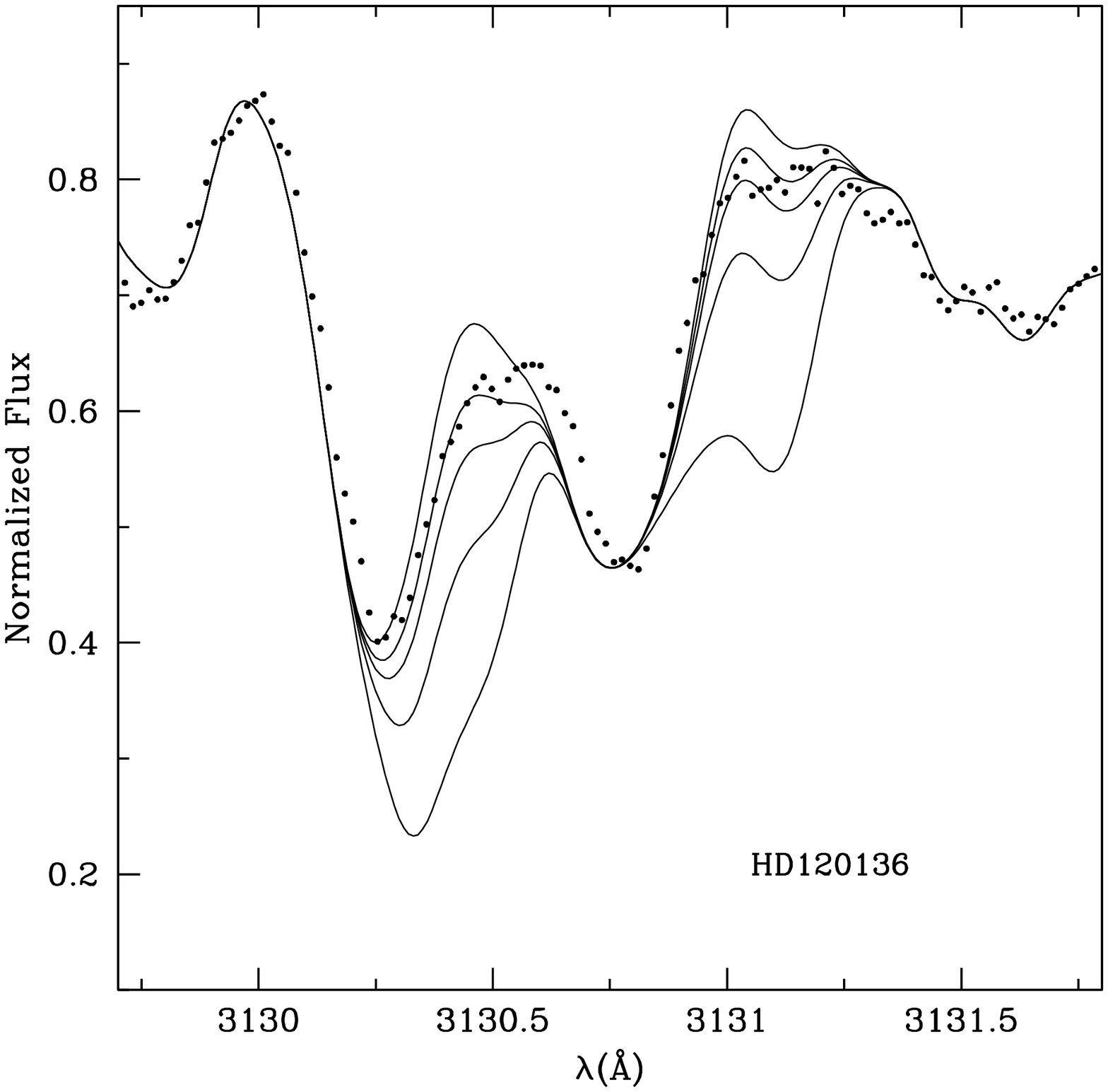}
\caption[]{Spectral syntheses (lines) and observed spectra (dots) in 
the \ion{Be}{ii} line region for the star \object{HD\,120136} ($\tau$\,Boo). The different curves 
correspond to syntheses with Be abundances of 1.42 (meteoritic), 0.65, 0.25, -0.05, 
and $-10.0$ (no Be).}
\label{figtauboo}
\end{figure}

\subsection{Spectral synthesis}

The abundance analysis was done in standard Local 
Thermodynamic Equilibrium (LTE) using the 2002 version of the code 
MOOG\footnote{The source code of MOOG2002 can be downloaded at 
http://verdi.as.utexas.edu/moog.html} \citep{Sne73}, and a grid of \citet{Kur93} 
ATLAS9 atmospheres. Be abundances were derived by fitting synthetic spectra to the 
data, using the same line list as in \citet{Gar98}. While both \ion{Be}{ii} 
lines at 3130.420 and 3131.065\,\AA\ are present in our data, we only used the latter, 
given the severe line blending in the region around 3130.420\,\AA\ 
(which has been used only for checking the consistency of the fit).

We derived the Be abundance for the Sun using a Kurucz model atmosphere \citep[][]{Kur93}
with T$_{\mathrm{eff}}$=5777\,K, $\log{g}$=4.44, and $\xi_{\mathrm{t}}$=1.0, and fitting 
the Kurucz Solar spectrum \citep[][]{Kur84}, after degrading its resolution to 70\,000,
a value similar to the resolution of the UVES spectra used in this paper. We
used a smoothing profile based on its $v\,\sin{i}$ \citep[1.9\,km\,s$^{-1}$ -- ][]{Sod82}, a
macroturbulence of 3.0\,km\,s$^{-1}$, and a limb darkening coefficient
of 0.6. The derived Be
abundance is $\log{N(Be)}$=1.10, only 0.05\,dex below the value obtained by \citet[][]{Chm75}. 
The small difference found is probably due to the different solar model
used by \citet{Gar98} to build the line-list. Since we are mostly interested in a
relative comparison, no changes were made in the line-list, and a solar value of
$\log{N(Be)}$=1.10 is considered in the rest of the paper.

When performing the spectral synthesis of the UES and UVES data, the synthetic spectra
were convolved with a Gaussian smoothing profile and a radial-tangential profile 
to take into account the spectral resolution and the macro-turbulence 
\citep[this latter was varied between 1.0 and 5.0 km\,s$^{-1}$, 
between K and F dwarfs - ][]{Gra92}, respectively.
A rotational profile was also added to account for the projected rotational velocity of the stars. 
The stellar $v\,\sin{i}$, listed in Tables\,\ref{tabplanets} and \ref{tabcomparison},
was for most of the cases estimated from the width of the CORALIE cross-correlation 
function \citep[see appendix of][]{San02b}. This method gives excellent results, 
as illustrated by \object{HD\,82943}, for which we have obtained 
the same value of $v\,\sin{i}$=1.65\,km\,s$^{-1}$ using the CORALIE CCF and a
detailed spectral synthesis method \citep[][]{Isr01}. A limb darkening coefficient
of 0.6 was considered for all cases, and the overall metallicity was 
scaled to the iron abundance. 

For the IACUB data only a Gaussian smoothing profile was added, to take
into account the instrumental profile. In this case, although the
spectrograph can provide R$\sim$50\,000 data, the observations were carried out at
lower resolution to improve its efficiency in the \ion{Be}{ii} spectral region
(as the instrument was attached to a telescope of 2.5-m diameter). At this
resolving power, the instrumental profile dominates the broadening of the
observed spectra (note the low $v\sin i$ of the targets).

We then iterated by changing the Be abundance and the continuum placement until the best 
fit for the whole spectral region was obtained. In this procedure, the global fit in 
the wavelength interval from 3129.5 to 3132\,\AA\ was considered, and not simply the 
region around the \ion{Be}{ii} lines. Small changes of the Gaussian smoothing profile 
were allowed when judged necessary (although these changes were always very small). 
The resulting abundances for all the objects observed are listed in 
Tables~\ref{tabplanets} and \ref{tabcomparison}. Here we use the 
notation $\log{N(Be)}$=[Be]=$\log{\mathrm{(Be/H})}$+12. Some examples of
fits are shown in Figs.\,\ref{figplanets} and \ref{figcomparison} (for 
more of such plots we refer to Paper\,B).

For \object{HD\,120136} ($\tau$\,Boo), the Be determination was particularly difficult 
since this F dwarf has a $v\,\sin{i}$ higher than all the other stars 
($\sim$15\,km\,s$^{-1}$). However, and as we can see from Fig.\,\ref{figtauboo}, from our
spectral synthesis we can be sure that this star has already depleted a good part of its Be.
Given the difficulty of the fit, however, the upper limit that we have determined for the 
Be abundance of $\tau$\,Boo ($+$0.25) should be considered as an approximate value. 

\subsection{Errors}

Measuring the uncertainties in the determination of Be abundances in not an easy 
task \citep[we refer to][ for a more thorough discussion]{Gar95}. In this 
paper we have adopted the following procedure. First of all, we considered that from 
the errors of $\pm$50\,K in temperature and $\pm$0.12\,dex in $\log{g}$ we can expect 
typical uncertainties around 0.03 and 0.05\,dex, respectively. An error of 0.05\,dex was 
further added to take into account the fact that there are several OH line-blends 
in the Be line region; uncertainties in the oxygen abundance will affect the location 
of the pseudo-continuum and introduce errors in the final Be abundance. 

These values were estimated by fitting the solar spectrum and varying the 
different parameters. This procedure also revealed that errors in other atmospheric parameters, 
like the metallicity and the microturbulence (with a derived uncertainty of the order 
of 0.05\,dex and 0.10\,km\,s$^{-1}$, respectively -- see Sect.\,\ref{sec:parameters}), do not influence the results significantly. Our experience also showed that the 
final Be abundance was slightly sensitive to the values for the broadening parameters 
used. We have thus added an extra 0.05\,dex uncertainty due to errors in the measured 
spectral resolution, macro-turbulence, and $v\,\sin{i}$.

Added quadratically, these figures produce an uncertainty of 0.09\,dex, which was 
added to the error due to continuum placement. This latter was estimated when fitting the spectrum 
and quantifies the quality of the fit. The final errors are of the order of 0.10-0.15\,dex; they 
are usually higher for the cooler dwarfs, and quite small for solar and hotter stars.

\citet[][]{Gar95b} studied in detail the sensitivity of the observed feature at $\lambda$\,3131\,\AA\ to the
Be abundance for low-mass stars. For effective temperatures below $\sim$5100\,K the feature
starts to be dominated by the contribution of another element (likely \ion{Mn}{i}). This makes it 
difficult, and sometimes impossible, to determine accurate Be abundances for the coolest
stars of our sample; the errors involved in the Be determinations are higher for
these stars. 

The usefulness of the \ion{Be}{ii} line for deriving Be abundances will thus 
decrease with temperature to a point where it will be useless. However, the 
exact T$_{\mathrm{eff}}$ where this will occur will depend on the resolution and signal-to-noise 
ratio of the observed spectra. And even if the observed feature is not dominated by the Be transition,
it is possible to obtain a reliable Be measurement by fitting synthetic
spectra with different Be abundances, as long as the S/N of the data is high enough.

In this sense, three low-mass stars studied (\object{HD\,22049} - 5073\,K, \object{HD\,26965\,A}
- 5126\,K, and \object{HD\,74576} - 5000\,K) do have spectra with signal-to-noise ratios high
enough to allow precise fitting and an abundance measurement. The analysis of
other objects with similar or cooler temperatures just provide upper limits,
reflecting both these difficulties and the decrease of Be abundances with
decreasing T$_{\mathrm{eff}}$ (see Sect.\,\ref{sec:be}). Examples of fits for these stars 
can be seen in Figs.\,\ref{figplanets} and \ref{figcomparison}: \object{HD\,22049} and 
\object{HD\,26965\,A}, with Be measurements, and \object{HD\,13445}, with just an upper limit.

\subsection{Instrumental offsets}
\label{sec:offset}

Given that our data were obtained with three different instruments, it is important 
to check for possible systematic errors in the Be abundances obtained from different
runs. This is particularly important since, for example, most of the IACUB spectra 
concern planet-host stars, while most of the UVES(C) data regard comparison sample
objects. If the systematic differences between the different instruments are significant, 
such a result could compromise a comparison between the planet hosts and the comparison 
sample stars.

\begin{figure}[t]
\psfig{width=\hsize,file=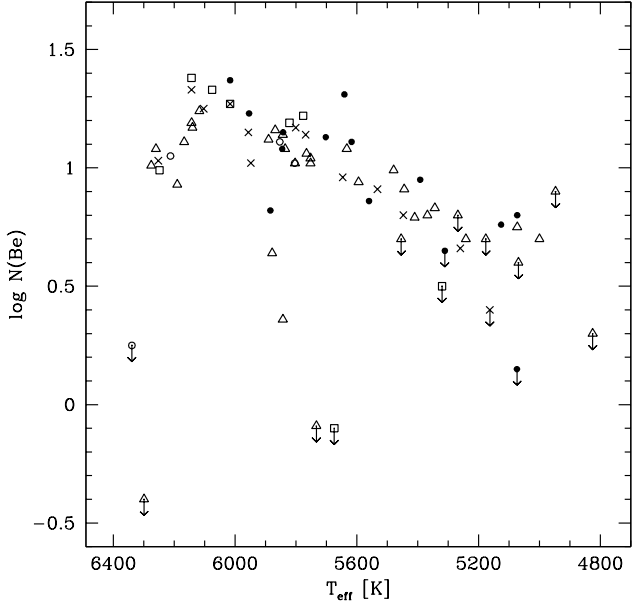}
\caption[]{Derived Be abundances as a function of effective temperature for
the stars listed in Tables\,\ref{tabplanets} and \ref{tabcomparison}. Crosses represent
results from UVES(A) spectra, open squares from UVES(B), open triangles from
UVES(C), open circles from UES, and dots represent Be abundances as derived from
IACUB spectra.}
\label{fig_instruments}
\end{figure}

Sources of systematic errors can be related, for example, to errors in the background 
correction during the reduction and extraction of the spectra. 
If the flux in the spectral region of interest is very
low, we might expect a more uncertain background correction.
Other errors might arise from the resolution achieved in the different spectra. 
The IACUB data used in this paper have a spectral resolution close to the limit as imposed by
the ability to resolve the \ion{Be}{ii} features as well as other lines in
the region. As discussed in \citet[][]{Gar95}, for spectra with resolution lower 
than $\sim$30\,000 the Be abundances can no longer be accurately obtained.
Furthermore, errors in the smoothing parameters used to fit the spectra (e.g. 
errors in the estimated spectral resolution itself) might also induce
considerable deviations in the final abundances; this is particularly true for
the lowest resolution spectra, where the instrumental profile dominates.

In Table\,\ref{tabplanets}, three of our target stars 
(\object{HD\,22049}, \object{HD\,75289}, and \object{HD\,82943}) have more than one
Be abundance measurement, obtained using spectra taken with different instruments, or
in different observing runs. This gives us the possibility of testing our results for systematic errors due to the different instrumental sets. 

In general, the values in the table seem to suggest that no major
systematic differences exist. A very small, and not significant difference ($\sim$0.05\,dex) 
is found between the $\log{N(Be)}$ derived from the IACUB and UVES, and no major 
conclusions can be drawn with the number of comparison points available. Note, however,
that the IACUB spectrum obtained for \object{HD\,82943} has a particularly low signal-to-noise 
ratio.

In Fig.\,\ref{fig_instruments} we present a plot of Be abundances for our
targets as a function of T$_{\mathrm{eff}}$. In the figure, the different symbols
indicate different instruments with which the spectra used to derive the Be abundances were obtained. This plot further attests that there are no significant differences between
the Be abundances derived from the different instruments.

\section{Li abundances}

The lithium abundances presented in Tables\,\ref{tabplanets} and \ref{tabcomparison}
are revised versions of those derived by \citet[][]{Isr04}. The new values 
were obtained using the ``new" stellar atmospheric parameters presented in \citet[][]{San04}, 
the same used to derive the Be abundances. 
The differences with the values listed by \citet[][]{Isr04} are always very small,
as expected, since the formerly used atmospheric parameters were not very different from 
the current ones. 

For a few stars \citet[][]{Isr04} had not derived Li abundances. These were now
computed in the same way as in their former work. We refer to this paper for details.
In the rest of this paper, the solar Li abundance was taken 
from \citet[][]{Gre98}, $\log{N(Li)}$=1.10.

\section{Be in planet-host stars}
\label{sec:be}

\begin{figure}[t]
\psfig{width=\hsize,file=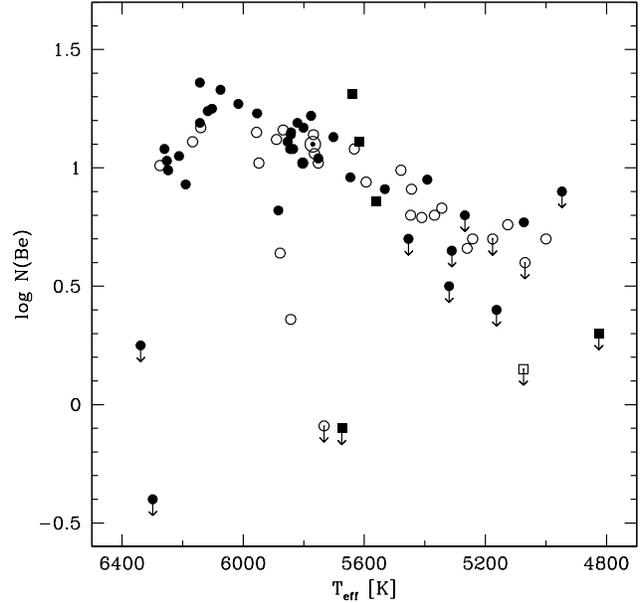}
\caption[]{Derived Be abundances as a function of effective temperature for
the stars listed in Tables\,\ref{tabplanets} and \ref{tabcomparison}. Planet hosts are
denoted by the filled symbols, while open symbols denote ``single'' stars. Circles represent
dwarfs, while the squares indicate sub-giants. The Sun is denoted by the usual symbol.}
\label{fig_be}
\end{figure}

In Fig.\,\ref{fig_be} we plot the derived Be abundances as a function of effective
temperature for our program stars. In the plot, the open symbols denote ``single'' stars,
while closed symbols represent planet hosts. 

In this figure, sub-giants are denoted with squares (see Paper\,B). The Be abundances 
for these stars should be taken with caution in the following 
comparison, since their convective envelopes are deepening. Our sample does not 
have many stars in this situation, and we prefer not to include these in most
of the discussion. The Be abundances of these objects will be further discussed in Paper\,B.

\begin{figure}[t]
\psfig{width=\hsize,file=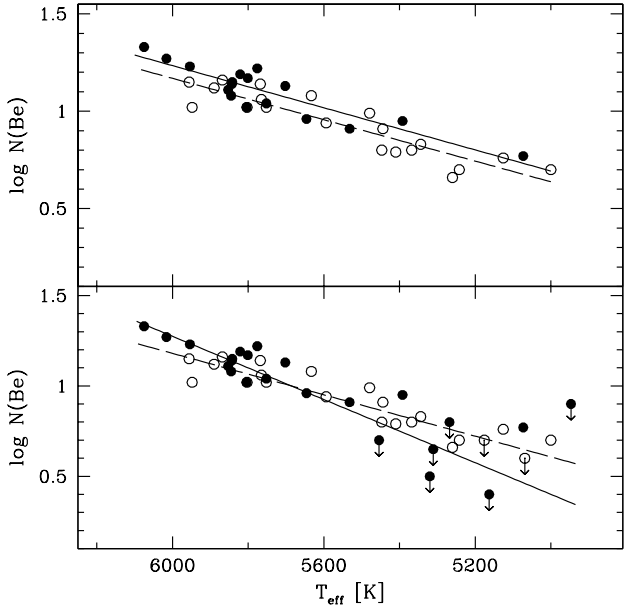}
\caption[]{Be abundances as a function of effective temperature for planet hosts (dots) and
comparison sample stars (open circles). Only dwarfs with T$_{\mathrm{eff}}$ 
below 6100\,K were considered. The 4 solar-temperature dwarfs that fall outside of the main
trend were not included wither (see text for more details). {\it Upper panel}: The two lines represents
linear fits to each of the two groups of stars (continuous line for planet-hosts, and
dashed line for comparison sample stars); only objects with Be detections were
considered in this plot. {\it Lower panel}: Same as above, but when upper limits are
also considered. The two lines represent fits to the data using a Buckey-James method (see
text for more details).}
\label{figfits}
\end{figure}

\begin{figure}[t]
\psfig{width=\hsize,file=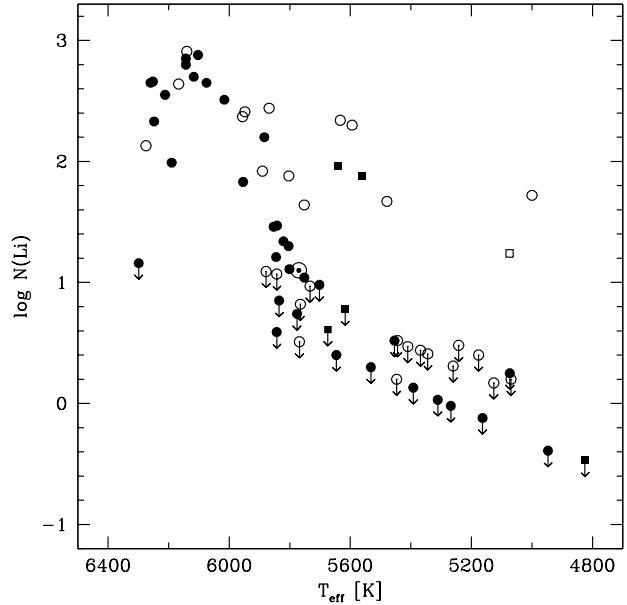}
\caption[]{Li abundances as a function of effective temperature for
the same stars as plotted in Fig.\,\ref{fig_be}. The Sun is denoted by the usual symbol.}
\label{figliteff}
\end{figure}

A visual inspection of Fig.\,\ref{fig_be} tells us that at a first sight, and overall, 
no clear difference seens to exist between the two populations of stars. 
Planet hosts and ``single'' stars describe the same trend in 
the $\log{N(Be)}$ vs. T$_{\mathrm{eff}}$ space. Globally, the Be abundances of both samples
decrease from a maximum near T$_{\mathrm{eff}}$=6100\,K, towards both higher
and lower temperature regimes (for a throughout discussion see Paper\,B).
This observation is in agreement with former results on the subject \citep[][]{San02},
and is similar to the trend observed for Li (see Fig.\,\ref{figliteff}).

In Fig.\,\ref{figfits} (upper panel) we further plot the Be abundances as a function of effective 
temperature for planet hosts (dots) and comparison sample dwarfs (open circles) with 
effective temperatures below 6100\,K. Only stars with Be determinations are 
considered (upper limits were not used).
The group of dwarfs with solar temperature that fall considerably below
the global trend were also excluded from this plot. As we discuss in Sect.\,\ref{anomalies}
and more thoroughly in Paper\,B, these stars seem to define a Be gap.
Two linear fits to the data are also shown,
the continuous line denoting a fit to the planet-host star points, and the dashed 
line a fit to the comparison sample star data.
As we can see from the figure, and although the visual difference between the two
groups of points is not very clear, there seems to be a systematic difference
in the sense that planet hosts are more Be-rich than comparison sample stars.
This difference, almost constant for all the temperatures, is of the order 
of 0.05\,dex. The slopes of the two fits are 
of 0.56$\pm$0.07\,dex/1000\,K (planet-hosts) and 0.53$\pm$0.07\,dex/1000\,K (comparison stars),
and the intercept values are $-$2.1$\pm$0.4\,dex (planets) and $-$2.0$\pm$0.3\,dex (comparison).
The rms of the fits is 0.06 and 0.07 for planet hosts and comparison sample stars, respectively.

If we take into account the upper-limit measurements,
the situation is a bit different. In Fig.\,\ref{figfits} (lower panel)
we present such a comparison. The two fits were now done with a
Buckey-James method, using ASURV Rev.\,1.4 \citep[][]{LaV92}, which implements the
methods presented in \citet[][]{Iso86}. The slopes and intercept values (slope,intercept) for the
two cases are (0.87$\pm$0.10,$-$3.97) for planet-hosts, and (0.57$\pm$0.06,$-$2.26) for
comparison stars.
These fits seem to suggest that for the lowest-temperature
stars in our sample, planet hosts may be a bit more Be-poor than
comparison sample stars. The difference in the slope of the two
samples is mostly due to the few upper-limit Be
abundances obtained for the planet hosts in the temperature regime
bellow $\sim$5500\,K, while for hotter dwarfs the difference is not as large.

In this sense, it has been shown by \citet[][]{Isr04} that the Li abundances
of planet-hosts are not above the ones found for field dwarfs. Instead,
there is some evidence that they are lower in the temperature regime between
5600 and 5850\,K \citep[see ][]{Isr04}. This fitting result could be hinting
at a similar result for Be in the lower temperature regime. 
However, the number of points available is probably not enough to
reach a definite conclusion regarding this matter.

Given the low number of measurements, these results do not seem to be
particularly significant. All the two-sample tests described in ASURV \citep[see][]{Fei85}
give a low probability that the two samples are not part of the same
distribution\footnote{This is true, whether if we fit and subtract the
trend of decreasing Be with decreasing temperature or not.}. Furthermore, it should be 
mentioned that the Buckey-James method to fit data containing non-detections must 
be taken with caution, since it assumes that the upper limits
in a given experiment are precisely known, while here these somehow represent
``n'' times the photon noise of the spectra.

\begin{figure}[b]
\psfig{width=\hsize,file=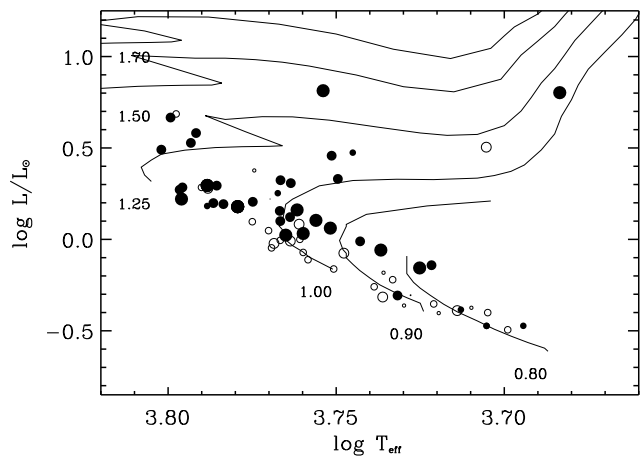}
\caption[]{HR diagram for the stars discussed in the current paper. Open circles
denote comparison sample stars, while planet hosts are shown as closed circles.
The size of the symbols is proportional to the stellar metallicity.   
Solar metallicity isochrones from \citet[][]{Lej01} for 0.8 to 1.7 M$_{\sun}$ are 
also shown.}
\label{fighr}
\end{figure}

On the other hand, if we ``believe'' in the upper panel of Fig.\,\ref{figfits},
what could possibly explain the apparent Be-enhancement in planet-host
stars? 

We can probably exclude that it is due to any pollution effects, since the
accretion of hydrogen-poor, metal-rich material would induce a larger variation for the hottest stars in Fig.\,\ref{figfits}, since these
have shallower convective envelopes. This is not
seen in the fits (if pollution was responsible for the observed difference, then this 
would also impose a strict limit on the quantity of accreted material). 
Curiously, a higher slope is found for planet hosts once we take into
account the upper-limit measurements (Fig.\,\ref{figfits}, lower panel).
But as mentioned above, the difference between the two fits (planet hosts and comparison)
in this case is mostly due to the lower temperature points, and not to
the hotter dwarfs, which would be more sensitive to pollution effects.

One better possibility is that the observed difference is due to galactic evolution effects. 
Planet hosts are metal-rich when compared with the ``single" stars in our sample. As
was shown e.g. by \citet[][]{Reb88}, \citet[][]{Mol97}, and \citet[][]{Boe99}, there 
is a trend for the Be abundances to increase with increasing metallicity, reflecting the Be evolution
of the galaxy. A multi-linear fit to the stars in the upper panel of Fig.\,\ref{figfits} gives
$\log{N(Be)}$=$-$2.00+0.53\,(T$_{\mathrm{eff}}$/1000)+0.12\,[Fe/H],
a relation suggesting that for a given temperature the
Be abundances increase with the metallicity (this point will be further discussed
in Garc\'{\i}a L\'opez et al., in preparation). 
Since NLTE effects on the Be abundances 
are not very strong \citep[][]{Gar95b}, they may not be the cause of this difference
for stars with different metallicity.

In Fig.\,\ref{fighr} we present an HR-diagram of the stars studied in this
paper. The size of the symbols is proportional to the stellar metallicity.
The luminosity was computed using Hipparcos parallaxes and V magnitudes \citep[][]{ESA97} , and
the bolometric correction of \citet[][]{Flo96}\footnote{Since the values in his tables are
wrong, we have derived the calibration of T$_{\mathrm{eff}}$ vs. BC by fitting the data 
in the paper in the same way as he did.}.
Planet hosts are denoted by open symbols, while comparison sample stars
are presented as closed symbols. As can be seen from the plot,
planet hosts seem to be, for a given temperature, a bit more luminous
than comparison sample stars. This difference may be related to the
higher average metallicity of the planet-host stars \citep[e.g.][]{San04},
or to a slighly different evolutionary status of these stars 
\citep[radial-velocity surveys for planets are more sensitive to older, non-active stars --][]{San00}. 
However, this difference also means that planet hosts are, for
a given temperature, a bit more massive than comparison sample objects.
Whether this mass ``excess'' could (also) be responsible for the eventual
difference observed in Fig.\,\ref{figfits} is not clear at this moment.

In general, these results argue against pollution as the key process leading to an overall 
metallicity excess of stars with planets \citep[see e.g. ][]{San03,San04,Pin01}. 
Even the 0.05\,dex difference observed in Fig.\,\ref{figfits} (upper panel), if due to any 
pollution events, would not suffice to explain the difference in the observed [Fe/H].
For example, adding $\sim$50 earth masses of C1 chondrites to the Sun would increase 
its iron abundance by about 0.25\,dex (a value similar to the average difference 
observed between stars with and without detected giant planets), and its Be 
abundance would increase by a slightly higher factor\footnote{For this calculation 
we have used a Be/Fe ratio similar to C1 chondrites, and considered an iron 
content of $\sim$20\% by mass \citep[][]{And89}.}. No difference of such magnitude 
seems present in our data. 

In the same way, the results do not support either 
extra mixing due to an eventual different angular momentum history of the 
two ``populations'' of stars. As mentioned above, \citet[][]{Isr04} (see their Fig.\,5) 
have shown that in the temperature regime between 5600 and 5850\,K, planet hosts
present lower Li abundances than do single field
dwarfs. This difference is not clear for Be (see Figs.\,\ref{fig_be} and \ref{figfits}), 
although the result shown in Fig.\,\ref{figfits} (lower panel) could hint at this being the case for lower temperature stars.

\subsection{Beryllium anomalies}
\label{anomalies}

A few solar temperature stars in Fig.\,\ref{fig_be} seem to show 
particularly low Be abundances. These objects, mostly ``single'' stars, seem to define
a Be-gap for temperatures between roughly 5600 and 6000\,K. At this moment we have no 
reasons to believe that these strange anomalies have anything to do with the presence (or not) 
of planets. However, more data are needed.

In \citet[][]{San02} we mentioned that contrary to what is expected from the
Be depletion models including rotational mixing \citep[e.g.][]{Pin90}, the Be abundances show 
a decreasing trend as a function of decreasing effective temperature. A few 
possible explanations for this trend 
were discussed, including the possibility that it has some relation
to the presence of planets around the targets.
At that time, however, the ``comparison''
stars with Be abundances available were completely outnumbered by the planet hosts. The current analysis has now overcome that problem,
and seems to show that the same trend is present for both planet-hosts
and ``single'' stars. In other words, the decreasing trend observed
in Fig.\,\ref{fig_be} does not seem to be related to the presence or not of
a planet orbiting the star, and is probably related to the Be depletion
mechanisms. It should be mentioned, however, 
that e.g. \citet[][]{Boe02} and \citet{Boe03} have not found the same trend in their studies of young 
open-cluster F and G dwarfs.

We refer to Paper\,B for a thorough discussion of these observed anomalies.

\begin{figure}[t]
\psfig{width=\hsize,file=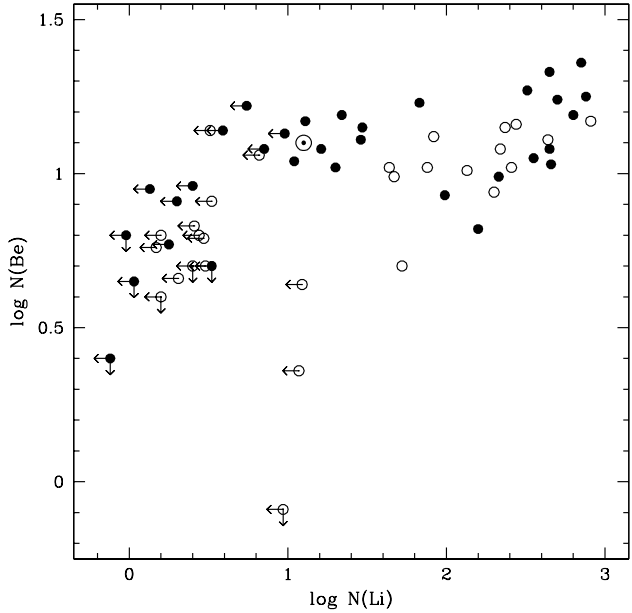}
\caption[]{Li and Be abundances plotted against each other. Only dwarfs were considered in
this plot. Closed circles represent planet hosts, while open symbols denote comparison
sample stars. The Sun is denoted by the usual symbol.}
\label{figlibe}
\end{figure}

\subsection{Lithium and Beryllium}

In Fig.\,\ref{figliteff} we present a plot of the Li abundances as a function of
effective temperature for the stars in our sample. This plot shows the usual
behavior of Li with temperature, and with only a few exceptions, 
no particular difference seems to exist between the two groups of 
stars \citep[see ][]{Isr04}. However, 4 objects in Fig.\,\ref{figliteff} 
(\object{HD\,36435}, \object{HD\,43162}, \object{HD\,43834}, and \object{HD\,74576}), 
all ``single" dwarfs, with effective temperatures below 5700\,K,
have particularly high Li abundances, clearly falling off the main 
trend. As we discuss in Paper\,B, however, and except for \object{HD\,43162}, their position 
in this plot is probably due to their young ages, and has probably nothing to do with the presence 
or not of a planet. However, the last mentioned star is particularly troubling, since
it is not clear if it is young. If its youth cannot explain the
high-Li content of \object{HD\,43162}, then other explanations have to be
considered. Either this star has recently engulfed planetary material (Li-rich
and H-poor), thus increasing considerably its Li content, or for some reason it has depleted
Li at lower rates than other dwarfs with a similar temperature. 

There are also three sub-giants in the above mentioned temperature regime 
(\object{HD\,10697} and \object{HD\,117176}, both planet hosts, and \object{HD\,23249}, 
a ``single" star), presenting observable Li abundances. The first two of these have been discussed in 
Sect.\,\ref{hintsofpollution} as well as in \citet[][]{San02}. All these 
stars will be further analyzed in Paper\,B. Amongst these three stars,
\object{HD\,23249} is the most interesting case, since its Be abundance
is clearly depleted, while it still has observable quantities of Li. 
This result might imply that ``pollution"
mechanisms have affected the photosphetic composition of this sub-giant.

A comparison of the Li and Be abundances for the dwarfs in our sample is
presented in Fig.\,\ref{figlibe}. A look at the figure suggests that globally 
there are no clear differences between planet hosts (closed circles), and ``single''
stars (open circles). In general, the stars follow the expected trend:
stars that are Be depleted are also severely Li depleted. There are, however, 
a few ``single'' stars with ``high''
values of Li and low values of Be. These cases correspond mostly to the Be-gap 
stars discussed in Sect.\,\ref{anomalies}, and their position in the plot has probably
nothing to do with the presence (or not) of planets. In Paper\,B
we discuss these points in more detail.

Curiously, there seems to be a lack of planet hosts with Li abundances 
($\log{N(Li)}$) between roughly 1.5 and 2.5\,dex, a region that is
populated by ``single'' stars. All planet hosts in this region
have, in general, lower Be abundances than do stars without known
planets. We note that this Li gap was also discussed in \citet[][]{Isr04}. Interestingly, 
the opposite effect seems to occur for Li abundances around $\log{N(Li)}$=1.0, a region for which
most of the targets are planet-hosts. We note that at least part
of these trends might be the result of the non-uniform temperature 
distribution of our two samples (planet hosts and comparison sample stars).

The small group of planet hosts with high Li and high Be
abundances described in Sect.\,\ref{hintsofpollution} (see below) is also clear in this
plot as the small group of points in the upper-right corner of the diagram.

\subsection{Outliers: Hints of ``pollution''?}
\label{hintsofpollution}

Interestingly, there is a group of planet-host stars with T$_{\mathrm{eff}}$ between
6000 and 6200\,K that present particularly high Be abundances. All these 
stars (\object{HD\,52265}, \object{HD\,75289}, \object{HD\,82943}, \object{HD\,121504}, 
and \object{HD\,209458}) have [Fe/H] above solar, in the range from 0.02 for \object{HD\,209458}
up to 0.30\,dex for \object{HD82943}, and Be abundances from 1.24 to 1.36\,dex. 
Given that their shallow convective envelopes are particularly easy to 
``pollute'' \citep[see e.g. discussion in ][]{Pin01}, this group of stars could 
represent a sample of objects that have seen their atmospheric abundances altered 
by the accretion of metal-rich, hydrogen-poor planetary material.

Of this group of objects, \object{HD82943} is probably the best candidate. In fact,
this star has already been found to have an anomalous abundance of $^6$Li, an indication
that it might have engulfed a planet, or at least planetary material \citep[][]{Isr01,Isr03}. 
The current result might thus confirm the former suspicions.

Unfortunately, the maximum in the Be abundances observed in this temperature range is 
very narrow, and we do not have many comparison stars in this temperature regime to verify
whether this result can be due to the presence of planets or not. Furthermore, 
on either side of this temperature interval (i.e. for T$_{\mathrm{eff}}>$6150 and 
T$_{\mathrm{eff}}<$6000), planet hosts and comparison sample stars do not show 
any remarkable difference. This peak in the Be abundances may simply correspond 
to a region for which Be depletion occurs at lower rate. 

Alternatively, and given that these objects are quite metal-rich, their initial 
Be abundances could have been already higher when compared with other 
stars \citep[][]{Reb88,Mol97,Boe99}, thus explaining their position in the 
plot (see also Sect.\,\ref{sec:be}).

A look at Fig.\,\ref{figliteff}, where we plot the Li abundances as a function of
effective temperature for the same stars plotted in Fig.\,\ref{fig_be} might also
give us some information about this point. In fact, these stars do not seem to
present particularly high Li abundances when compared with e.g. the stars without planets
that are present in the same temperature regime, although a peak in the Li 
abundances also appears at this temperature.

A sixth star may also deserve a comment. \object{HD\,10697}, a sub-giant with a temperature of
5640\,K, seems to have a particularly high Be abundance. This case was already discussed in
\citet[][]{San02}, as the sub-giant that together with \object{HD\,117176} presented 
a Li abundance that was high for its temperature. Given the uncertainties in the derived 
abundances, due e.g. to uncertainties in the surface gravity \citep[see][]{Gar98},
and to the use of a relatively low-resolution IACUB spectrum, 
and given that there may exist non-uniformities in the initial Be content of the stars, it is
not clear if this object is exceptionally Be-rich. If so, however, this could
mean one of various things: either this is a case where planetary material has been
engulfed, or the excess Be is due to a dredge-up effect from a ``buffer" below the former
main-sequence convective envelope \citep[][]{Del90}. It should be mentioned, however, that
this latter scenario does not seem to be supported by current observations \citep[e.g.][]{Ran99}.
The cases for the other sub-giants are discussed in Paper\,B.

\section{Concluding remarks}

In this paper we have obtained Be abundances for a set of 41 planet-host
stars, and a smaller sample of 29 stars not known to harbor any
planetary-mass companion. The abundances were derived from a detailed
spectroscopic analysis, and gave us the possibility to look for
possible differences between the two samples. 

A comparison of the Be abundances of planet hosts and ``single'' stars
has revealed that, perhaps with a few exceptions, the two samples
follow the same behavior in the $\log{N(Be)}$ vs. T$_{\mathrm{eff}}$
plot. A small offset of 0.05\,dex in Be abundance might be present for all
temperatures, planet hosts being more Be-riche. This small difference
is tentatively explained as due to the galactic chemical evolution.

The results presented support the idea that the excess metallicity
observed for planet hosts has, overall, a ``primordial'' origin, and is not
due to generalized stellar pollution processes. Nevertheless, we have found
a small group of stars that present particularly high Be abundances 
that could be explained by pollution events, although other explanations
are possible. More data are needed to allow us to take any conclusion.

In a separate paper (Paper\,B) we further analyse our results in the framework of
the models of Be depletion in solar-type stars.

\begin{acknowledgements}
  We wish to thank the Swiss National Science Foundation (Swiss NSF) 
  for the continuous support for this project. Support from Funda\c{c}\~ao 
  para a Ci\^encia e Tecnologia (Portugal) to N.C.S. in the form of 
  a scholarship is gratefully acknowledged.
\end{acknowledgements}


\begin{thebibliography}{}

\bibitem[Anders \& Grevesse(1989)]{And89}
Anders, R., \& Grevesse, N., 1989, Geochim. Cosmochim. Acta, 53, 197 

\bibitem[Barnes et al.(2001)]{Bar01}
Barnes, S., Sofia, S., \& Pinsonneault, M., 2001, ApJ, 548, 1071

\bibitem[Bodaghee et al.(2003)]{Bod03}
Bodaghee, A., Santos, N.C., Israelian, G., \& Mayor, M., 2003, A\&A, 404, 715

\bibitem[Boesgaard et al.(2003)]{Boe03}
Boesgaard, A.M., Armengaud, E., \& King, J.R., 2003, ApJ, 582, 410

\bibitem[Boesgaard \& King(2002)]{Boe02}
Boesgaard, A.M., \& King, J.R., 2002, ApJ, 565, 587

\bibitem[Boesgaard et al.(1999)]{Boe99}
Boesgaard, A.M., Deliyannis, C.P., King, J.R., Ryan, S.G., Vogt, S.S., Beers, T.C., AJ, 117, 1549

\bibitem[Chmielewski et al.(1975)]{Chm75}
Chmielewski, Y., M\"uller, E.A., \& Brault, J.W., 1975, A\&A, 42, 37

\bibitem[Deliyannis et al.(2000)]{Del00}
Deliyannis, C.P., Cunha, K., King, J.R., \& Boesgaard, A.M., 2000, AJ, 119, 2437S

\bibitem[Deliyannis et al.(1990)]{Del90}
Deliyannis, C.P., Demarque, P., \& Kalawer, S., 1990, ApJS, 73, 21

\bibitem[Ecuvillon et al. (2004)]{Ecu04}
Ecuvillon, A., Israelian, G., Santos, N.C., Mayor, M., Garcia Lopez, R., \& Randich, S., 2004, A\&A, 418, 703

\bibitem[Edwards et al.(1993)]{Edw93}
Edwards, S., Strom, S.E., Hartigan, P., et al., 1993, AJ, 106, 372

\bibitem[ESA(1997)]{ESA97}
ESA, 1997, The Hipparcos and Tycho catalogue, ESA-SP 1200

\bibitem[Feigelson \& Nelson(1985)]{Fei85}
Feigelson, E.D., \& Nelson, P.I., 1985, ApJ, 293, 192


\bibitem[Flower(1996)]{Flo96}
Flower, P.J., 1996, ApJ, 469, 355

\bibitem[Garc\'{\i}a L\'opez \& P\'erez de Taoro(1998)]{Gar98}
Garc\'{\i}a L\'opez, R.J., \& P\'erez de Taoro, M.R., 1998, A\&A, 334, 599 

\bibitem[Garc\'{\i}a L\'opez et al.(1995a)]{Gar95}
Garc\'{\i}a L\'opez, R.J., Severino, G., \& Gomez, M.T., 1995a, A\&A, 297, 787

\bibitem[Garc\'{\i}a L\'opez et al.(1995b)]{Gar95b}
Garc\'{\i}a L\'opez R.J., Rebolo, R., \& Perez de Taoro, M.R., 1995b, A\&A, 302, 184

\bibitem[Gonzalez et al.(2001)]{Gon01} 
Gonzalez G., Laws C., Tyagi S., \& Reddy B.E., 2001, AJ, 121, 432

\bibitem[Gonzalez \& Laws(2000)]{Gon00}
Gonzalez, G., \& Laws, C., 2000, ApJ, 119, 390

\bibitem[Gonzalez(1998)]{Gon98}
Gonzalez, G., 1998, A\&A, 334, 221

\bibitem[Gonzalez(1997)]{Gon97}
Gonzalez, G., 1997, MNRAS, 285, 403

\bibitem[Goodman \& Rafikov(2001)]{Goo01}
Goodman, J., \& Rafikov, R. R., 2001, ApJ, 552, 793

\bibitem[Gray(1992)]{Gra92}
Gray, D., 1992. In: ``The observation and analysis of stellar photospheres´´, Cambridge Univ. Press 

\bibitem[Grevesse \& Sauval(1998)]{Gre98}
Grevesse, N., \& Sauval, A.J., 1998, Space Sci. Rev., 85, 161

\bibitem[Hartmann(2002)]{Har02}
Hartmann, L., 2002, ApJ, 566, L29

\bibitem[Isobe et al.(1986)]{Iso86}
Isobe, T., Feigelson, E.D., \& Nelson, P.I., 1986, ApJ, 306, 490

\bibitem[Israelian et al.(2004)]{Isr04}
Israelian, G., Santos, N.C., Mayor, M., \& Rebolo, R., 2004, A\&A, 414, 601 

\bibitem[Israelian(2003)]{Israelian03}
Israelian, G., 2003, in IAU S219: Stars as Suns: Activity, Evolution, and Planets, ed. A.K. Dupree (San Francisco: ASP), in press (astro-ph/0310377)

\bibitem[Israelian et al.(2003)]{Isr03}
Israelian, G., Santos, N.C., Mayor, M., \& Rebolo, R., 2003, A\&A, 405, 753 

\bibitem[Israelian et al.(2001)]{Isr01}
Israelian, G., Santos, N.C., Mayor, M., \& Rebolo, R., 2001, Nature, 411, 163

\bibitem[Kurucz(1993)]{Kur93} 
Kurucz, R. L., 1993, CD-ROMs, ATLAS9 Stellar
Atmospheres Programs and 2~${\rm km}~{\rm s}^{-1}$ Grid
(Cambridge: Smithsonian Astrophys. Obs.)

\bibitem[Kurucz et al.(1984)]{Kur84}
Kurucz, R. L., Furenlid, I., Brault, J., Testerman, L., 1984, Solar Flux Atlas 
from 296 to 1300 nm, NOAO Atlas No.\,1

\bibitem[Molaro et al.(1997)]{Mol97}
Molaro, P., Bonifacio, P., Castelli, F., \& Pasquini, L., 1997, A\&A, 319, 593

\bibitem[LaValley et al.(1992)]{LaV92}
LaValley, M., Isobe, T., \& Feigelson, E.D., 1992, BAAS, Rev. 1.1

\bibitem[Laughlin \& Adams(1997)]{Lau97}
Laughlin, G., \& Adams, F.C., 1997. ApJ, 491, L51

\bibitem[Laws et al.(2003)]{Law03}
Laws, C., Gonzalez, G., Walker, K.M., et al., 2003, AJ, 125, 2664

\bibitem[Laws \& Gonzalez(2001)]{Law01}
Laws, C., \& Gonzalez, G., 2001, ApJ, 553, 405

\bibitem[Lejeune \& Schaerer(2001)]{Lej01}
Lejeune, T., \& Schaerer, D., 2001, A\&A, 366, 538

\bibitem[Murray \& Chaboyer(2002)]{Mur02}
Murray, N., \& Chaboyer, B., 2002, ApJ 566, 442

\bibitem[Naef et al.(2004)]{Nae04}
Naef, D., Mayor, M., Beuzit, J. L., Perrier, C., Queloz, D., et al., 2004, A\&A, 414, 351

\bibitem[Pinsonneault et al.(2001)]{Pin01} 
Pinsonneault, M.H., DePoy, D.L., \& Coffee, M. 2001, ApJ 556, L59

\bibitem[Pinsonneault et al.(1990)]{Pin90}
Pinsonneault, M.H., Kawaler, S.D., \& Demarque, P., 1990, ApJS, 74, 501

\bibitem[Randich et al.(1999)]{Ran99}
Randich, S., Gratton, R., Pallavicini, R., Pasquini, L., \& Carretta, E., 1999, A\&A, 348, 487

\bibitem[Rebolo et al.(1988)]{Reb88}
Rebolo, R., Molaro, P., Abia, C., \& Beckman, J.E., 1988, A\&A, 193, 193

\bibitem[Rebull(2001)]{Reb01}
Rebull, L.M., 2001, AJ, 121, 1676

\bibitem[Rebull(2002)]{Reb02}
Rebull, L.M., Wolff, S.C., Strom, S.E., \& Makidon, R.B., 2002, AJ, 124, 546

\bibitem[Reddy et al.(2003)]{Red03}
Reddy, B., Lambert, D., Laws, C., Gonzalez, G., \& Covey, K., 2003, MNRAS, 335, 1005

\bibitem[Reid(2002)]{Rei02}
Reid, I.N., 2002, PASP, 114, 306

\bibitem[Ryan(2000)]{Rya00}
Ryan, S., 2000, MNRAS, 316, L35

\bibitem[Sadakane et al.(2002)]{Sad02}
Sadakane, K., Ohkubo, M., Takeda, Y., et al., 2002, PASJ, 54, 911

\bibitem[Santos et al.(2004b)]{paperb} 
Santos, N.C., Israelian, G., Randich, S., Garc\'{\i}a L\'opez, R.J., \& Rebolo, R., 2004b, A\&A, in press (Paper\,B)

\bibitem[Santos et al.(2004a)]{San04} 
Santos, N.C., Israelian, G., \& Mayor, M., 2004a, A\&A, 415, 1153 

\bibitem[Santos et al.(2003a)]{sydney03} 
Santos, N.C., Mayor, M., Udry, S., et al., 2003a, in IAU S219: Stars as Suns: Activity, Evolution, and Planets, ed. A.K. Dupree (San Francisco: ASP), in press 

\bibitem[Santos et al.(2003b)]{San03} 
Santos, N.C., Israelian, G., Mayor, M., Rebolo, R., \& Udry, S., 2003b, A\&A, 398, 363

\bibitem[Santos et al.(2002a)]{San02}
Santos, N.C., Garc\'{\i}a L\'opez, R.J., Israelian, G., et al., 2002a, A\&A, 386, 1028

\bibitem[Santos et al.(2002b)]{San02b}
Santos, N.C., Mayor, M., Naef, D., et al., 2002b, A\&A, 392, 215

\bibitem[Santos et al.(2001)]{San01} 
Santos, N.C., Israelian, G., \& Mayor, M., 2001a, A\&A, 373, 1019

\bibitem[Santos et al.(2000a)]{San00} 
Santos N.C., Israelian G., \& Mayor M., 2000a, A\&A, 363, 228



\bibitem[Sari \& Goldreich(2004)]{Sar04}
Sari, R., \& Goldreich, P., 2004, ApJL, 606, 77

\bibitem[Siess \& Livio(1999)]{Sie99}
Siess, Lionel; Livio, Mario, 1999, MNRAS, 308, 1133

\bibitem[Smith et al.(2001)]{Smi01}
Smith, V.V., Cunha, K., \& Lazzaro, D., 2001, AJ, 121, 3207

\bibitem[Sneden(1973)]{Sne73} 
Sneden, C., 1973, Ph.D. thesis, University of Texas

\bibitem[Soderblom(1982)]{Sod82}
Soderblom, D.R., 1982, ApJ, 263, 269

\bibitem[Stassun et al.(1999)]{Sta99}
Stassun, K.G., Mathieu, R.D., Mazeh, T., \& Vrba, F.J, 1999, AJ, 117, 2941

\bibitem[Stephens et al.(1997)]{Ste97}
Stephens, A., Boesgaard, A.M., King, J.R., \& Deliyannis, C.P., 1997, ApJ, 491, 339

\bibitem[Strom(1994)]{Str94}
Strom, S.E., 1994. In: Jean-Pierre Caillault (ed.), ``8th workshop on Cool Stars, 
Stellar Systems, and the Sun'', ASP conf. series, 64, p.211

\bibitem[Vauclair(2004)]{Vau04}
Vauclair, S., 2004, ApJ, 605, 874

\bibitem[Wolff et al.(2004)]{Wol04}
Wolff, S.C., Strom, S.E., \& Hillenbrand, L.A., 2004, ApJ, 601, 979

\end{thebibliography}
\end{document}